\documentclass{JHEP}
\usepackage{amssymb,amsmath}
\newcommand{\ii}{\mathrm{i}}
\newcommand{\na}{\nabla}
\newcommand{\dd}{\mathrm{d}}
\newcommand{\pd}{\partial}
\newcommand{\hh}{\mathcal{H}}

\newcommand{\F}{\mathcal{F}}

\newcommand{\e}{\mathrm{e}}
\newcommand{\ket}[1]{\left|#1\right\rangle}
\newcommand{\bra}[1]{\left\langle #1\right|}

\newcommand{\tr}{\mathop{\mathrm{tr}}\nolimits}

\newcommand{\SW}{\mathop{\mathrm{SW}}\nolimits_{21}}

\newcommand{\I}{\mathbb{I}}
\newcommand{\V}{\mathcal{V}}
\title{Duality in non-commutative gauge theories as a non-perturbative
Seiberg--Witten map}

\author{Elias Kiritsis\\
Physics Department, P.O. Box 2208, University of Crete, and FORTH\\
GR-71003, Heraklion, GREECE\\ \email{kiritsis@physics.uoc.gr}}
\author{Corneliu Sochichiu\\ Bogoliubov Laboratory
of Theoretical Physics, Joint Institute for Nuclear Research,
141980 Dubna, Moscow Reg., RUSSIA \\ Institutul de Fizic\u a
Aplicat\u a A\c S, str. Academiei, nr. 5, Chi\c sin\u au, MD2028
MOLDOVA\\ \email{sochichi@thsun1.jinr.ru}}

 \preprint{hep-th/0202065}
\abstract{
 We study the equivalence/duality between various
 non-commutative gauge  models at the  classical and quantum
 level. The duality is realised by  a linear
 Seiberg--Witten-like map. The infinitesimal form of this map is analysed
 in more details.
 }

\begin{document}

\section{Introduction}

Quantum field theories over non-commutative spaces (for a review
see \cite{Connes:2000by}) seem to be relevant to the
non-perturbative description of the string theory
\cite{Ho:1997jr}\nocite{Ho:1998yk,Chu:1998qz,Connes:1998cr}-%
\cite{Seiberg:1999vs}.

Moreover, the non-commutative gauge models mimic some features
expected in the string field theory introduced by E.~Witten
\cite{Witten:1986cc}.

Thus, classical localised solutions called non-commutative solitons
found in the framework of non-commutative theory
\cite{Gopakumar:2000zd}\nocite{Polychronakos:2000zm,Sochichiu:2000rm}-%
\cite{Aganagic:2000mh}
in the string picture were shown to correspond to branes
\cite{Harvey:2000jt,Harvey:2000jb}.

In the non-commutative case, one can formally develop a perturbation
expansion similar to the commutative one. It was found that a
phenomenon called IR/UV mixing occurs due to non-commutativity
\cite{Filk:dm}\nocite{VanRaamsdonk:2000rr}--\cite{Minwalla:1999px}. This
again reflects a  feature of open/closed duality in string theory.  In
the case of non-commutative field theories it prevents us from giving
a direct proof of renormalisability of such theories.\footnote{An
exception may occure, however, when in commutative theory no
UV counterterms are needed at least at one loop e.g., like in
Wess--Zumino model.  In this case the IR/UV mixing is trivial
and no problems with renormalisability arise \cite{Girotti:2000gc}.}

On the other hand, string theory possesses a number of
\emph{dualities}  relating various string backgrounds
\cite{Witten:1995ex} (for reviews see  \cite{duality}),
\emph{M-theory\/} being the name of the embracing non-perturbative
model. In this context one naturally may ask if there are similar
duality relations in the non-commutative field theory models. Indeed,
in \cite{Connes:1998cr,Schwarz:1998qj} it was found that gauge models
on different but dual non-commutative tori are the same due to
\emph{Morita equivalence\/}.

In the case of non-commutative planes, it was shown in a
$p$-dimensional non-commutative gauge model one can recover a set of
$p\pm 2k$ dimensional gauge models as expansions around different
non-perturbative solutions in the above model
\cite{Sochichiu:2000bg,Sochichiu:2001am}. Depending on whether $p$ is
even or odd the models belong to the universality class of $N=\infty$
IKKT \cite{Ishibashi:1996xs}, or BFSS \cite{Banks:1997vh} M(atrix)
models. The above matrix models arise as reductions of ordinary
SU($N$) Yang--Mills model to, respectively, zero and one
dimension. Such a relation between large $N$ reduced model and the
noncommutative model appears to be an avatar of the old idea by
Eguchi--Kawai (see \cite{Eguchi:1982nm}).

{}From a different point of view, the above map between different
models can be interpreted as a non-perturbative Seiberg--Witten map
since it is an explicit solution to the Seiberg--Witten equation
\cite{Seiberg:1999vs}, which depends non-perturbatively on the gauge
coupling and the non-commutativity parameter.

Traditionally, attempts to solve and classify the solutions to
Seiberg--Witten equations were made using the power expansion of
non-commutativity parameter $\theta$ originally introduced in
\cite{Seiberg:1999vs}. Although, considerable progress was achieved
since that time \cite{Dimakis:2000vm}--%
\nocite{Jurco:2001rq,Bichl:2001nf,Brace:2001rd,Asakawa:1999cu}%
\cite{Goto:2000zj}, this approach remains extremely complicated.

Here we propose an alternative way to find the Seiberg--Witten map
starting from background independence. The background independent
formulation in our context is the formulation of the non-commutative
gauge model in terms of Hermitian (or as appropriately required)
operators acting on a separable infinite-dimensional Hilbert space
with no explicit dependence on the space coordinates or
derivatives. The non-commutative spaces in this approach appear as
classical background solutions to these models, which break
(spontaneously) the background invariance.  Seiberg--Witten map in
this picture relates formulations of the model around different
backgrounds. The particular property of the map we describe is that it
is linear in fields.

In the present work we explore the duality realised by the map both in
the non-perturbative case, when it relates the models in
non-commutative spaces of different dimensionality
\cite{Sochichiu:2000bg,Sochichiu:2000kz}, or models with different
gauge groups, and in the perturbative case it corresponds to a smooth
change of the non-commutativity parameter \cite{Polychronakos:2000nt}.
We extend the analysis to the quantum case when duality is realised as
a quantum symmetry in the path integral approach.

The plan of the paper is as follows.  In section 2 we review the
non-commutative gauge field theory. We introduce the gauge model of
Yang-Mills interacting with a Higgs scalar multiplet as an expansion
of the the bosonic part of $N=\infty$ matrix model around a classical
solution of the model.\footnote{This approach is similar in its main
lines to the one used in \cite{ikkt}. Note, however, the difference in
notations for the gauge field.}. In section 3 we describe the
dualities in the non-commutative gauge model first in their most
general form and afterwards specialising to particular cases of
dualities relating models with different gauge groups as well as
relating models in different dimensions. In Section 4 we focus on the
infinitesimal form of proposed transformation which appears to be a
linear variant of the Seiberg--Witten map. Further, in Section 5 we
analyse quantum implications of the symmetry realised by the map.
Finally section six contains our conclusions and further remarks.

\section{The Model}

Large $N$ BFSS and IKKT matrix models were proposed to describe
non-perturbative string theory
\cite{Ishibashi:1996xs,Banks:1997vh}. They are reductions of the ten
dimensional SU$(N)$ Yang--Mills (YM) models to respectively zero and
one dimensions. On the other hand, it can be shown that
infinite-dimensional Hilbert space ($N=\infty$) models are again
YM-like, but living either on commutative or  non-commutative spaces.
In the limit $N\to\infty$, such models were shown to have the
infinite-dimensional gauge group of area preserving diffeomorphisms
\cite{Sochichiu:2000ud}.

In spite of an apparent dissimilarity, these models are related to the
``usual'' non-commutative U(1) Yang--Mills model by an appropriate
redefinition of the gauge fields.

Here, we will analyse the situation where Euclidean ``space-time''
coordinates are all non-commutative,
\begin{equation}\label{nc}
  [x^\mu,x^\nu]=\ii \theta^{\mu\nu},\qquad \det\|\theta^{\mu\nu}\|\neq
0.
\end{equation}
This can be easily generalised to degenerate $\theta^{\mu\nu}$. In
such a context,  the commutative variable are treated as parameters.

The model we consider here corresponds to (the bosonic part of) the
IKKT model. This model is given by the background invariant action of
the following form,
\begin{equation}\label{action}
  S=-\frac{1}{4g^{\prime 2}}\tr[X_i,X_j]^2,
\end{equation}
where the gauge fields $X_i$ are Hermitian operators acting on the
Hilbert space $\hh$. In this form, the model is formulated entirely in
terms of an abstract Hilbert space and Hermitian operators acting on
it. It contains no explicit space-time data which we introduce later
as particular solutions breaking this background invariance.

Equations of motion corresponding to the action \eqref{action} look as
follows,
\begin{equation}\label{eqm}
  [X_i,[X_i,X_j]]=0.
\end{equation}

Picking up a particular solution to the equations of motion in the
form $X_i^{(0)}=\Lambda_i^\mu p_\mu$, where $p_\mu$ is a complete
irreducible set of operators, (i.e. ones satisfying,
\begin{equation}\label{irred}
  [p_\mu,p_\nu]=-\ii \theta^{-1}_{\mu\nu},\qquad\forall F:
  [p_i,F]\Rightarrow F\sim\I,
\end{equation}
where the last expression is Schur's lemma implying the irreducibility
of the representation of the algebra generated by $p_\mu$), one can
expand operators $X_i$ around this background as follows
\begin{equation}\label{expantion}
  X_i=\Lambda_i^\mu (p_\mu+A_\mu)+\Phi_i,
\end{equation}
where $\Phi_i\Lambda_i^\mu=0$.

By an appropriate linear transformation depending on $\Lambda_i^\mu$
one can make $X_i$ to acquire the following form,
\begin{equation}\label{base-sol}
  X_\mu=p_\mu+A_\mu,\quad\mu=1,\dots,p;\quad X_a=\Phi_a,\quad
  a=p+1,\dots,D.
\end{equation}

Now, taking Weyl ordering with respect to operators
$x^\mu=-\theta^{\mu\nu}p_\nu$ will mean that in the chosen background,
the model describes a gauge field $A_\mu(x)$ interacting with a
``multiplet'' of scalars $\Phi_a(x)$.

Indeed, the Weyl transformation maps an operator $\phi$ to its symbol
according to the following rule
\begin{equation}\label{Weyl}
  \phi (x)=\sqrt{\det\theta}\int \frac{\dd^p k}{(2\pi)^{p/2}} \e^{\ii
  k\cdot x} \tr \e^{\ii k\times p} \phi,
\end{equation}
where $k\times p=k_\mu\theta^{\mu\nu}p_\nu$ and $k\cdot x=k_\mu
x^\mu$. Under this map, the operator products are transformed into
star products of symbols given by,
\begin{equation}\label{star}
  \phi*\chi(x)=\e^{\frac{\ii}{2}\theta^{\mu\nu}\pd_\mu\pd'_\nu}
  \phi(x)\chi(x')|_{x'=x},
\end{equation}
where $\pd'_\mu$ denotes derivatives with respect to $x^{\prime
\mu}$. The derivatives can be expressed algebraically in terms of the
star product as well,
\begin{equation}\label{star-drv}
  \pd_\mu \phi=\ii[p_\mu,\phi]_*(x),\qquad \na_\mu \phi=\ii
  [X_\mu,\phi]_*(x)=\ii [(p_\mu+A_\mu),\phi]_*(x).
\end{equation}

The Weyl map \eqref{Weyl} is invertible. The inverse is given by
\begin{equation}\label{invWeyl}
  \phi = \int\frac{\dd^pk}{(2\pi)^p}\dd x\,\e^{-\ii k\times p-\ii
  k\cdot x}\phi (x).
\end{equation}

The action can be written as a functional over non-commutative
functions,
\begin{equation}\label{p-dim}
  S=\int \dd^p x\ \left(-\frac{1}{4g^2}(\F_{\mu\nu}
-\theta_{\mu\nu}^{-1})^2 +\frac{1}{2g^2}\na_\mu\Phi_a*\na_\mu\Phi_a-
\frac{1}{4g^2}[\Phi_a,\Phi_b]_*^2\right),
\end{equation}
where,
\begin{subequations}\label{defs}
  \begin{align}\label{F} &\F_{\mu\nu}=\pd_\mu A_\nu-\pd_\nu
  A_\mu+[A_\mu,A_\nu]_*\\ \label{covariant}
  &\na_\mu\Phi=\pd_\mu\Phi+[A_\mu,\Phi]_*, \end{align} and the gauge
  coupling $g^2$ is defined by,
\begin{equation}\label{g}
  g^2=g^{\prime 2}\sqrt{\det\theta}.
\end{equation}
\end{subequations}
The products in eq. \eqref{p-dim} should be taken as star products,
however, one star product in each monomial can be substituted by the
usual product following the identity,
\begin{equation}
  \int\dd^p x\, \phi*\chi=\int\dd^px\, \phi(x)\chi(x).
\end{equation}

Let us note that the system possesses a ``local'' U(1) gauge symmetry
which in the background invariant formulation is the unitary symmetry
of the Hilbert space.

Another difference from the ``standard'' Yang--Mills model which can
be observed in the gauge field part of the action \eqref{p-dim} is
that the field strength $\F_{\mu\nu}$ comes shifted by the quantity
$\theta^{-1}_{\mu\nu}$. Although, for constant $\theta^{-1}_{\mu\nu}$
this shift has no effect on the equations of motion, it is in
conformity with the presumed string theory origin of the model, where
the gauge field appears in combination $\F_{\mu\nu}+B_{\mu\nu}$ with
the stringy $B_{\mu\nu}$-field.

Two comments are in order:
\begin{itemize}
 \item If in the solution, the set of operators $p_\mu$ is reducible
 in the sense that the second condition of \eqref{irred} is not
 fulfilled, one should complete the set by appropriate operators to
 make it irreducible. Operators $q_\alpha$ $\alpha=p+1,\dots,p'$
 commuting with $p_\mu$,
\begin{equation}\label{qa}
  [p_\mu,q_\alpha]=0,
\end{equation}
form a closed algebra,
\begin{equation}\label{cl-alg}
  [q_\alpha,q_\beta]=\ii c_{\alpha\beta}(q).
\end{equation}
In particular, this can be a finite dimensional Lie algebra or another
piece of the Heisenberg algebra. In the last case,  we have new
additional non-commutative coordinates which the fields should depend
on, while in the first case one has a non-abelian gauge symmetry
corresponding to representations of the algebra \eqref{cl-alg}. We
don't know to which extent the representation of the algebra
\eqref{qa} can depend on the space ``points''
$x^\mu=-\theta^{\mu\nu}p_\nu$ in general, but the simplest case is
when the Hilbert space is split as $\hh'\otimes V$, where $\hh'$ is
the infinite dimensional subspace of the Hilbert space on which
restriction of $p_\mu$ generate an irreducible representation while
$V$ realises an irreducible representation of the algebra \eqref{qa}.

For example, for $p_\mu=\I_{(n)}\otimes p'_\mu$, where $p'_\mu$ form
an irreducible set and $\I_{(n)}$ is the $n\times n$ unity matrix one
has the set of matrices $\sigma_{\alpha}$, $\alpha=1,\dots,n^2-1$
commuting with $p_\mu$. In this case, an arbitrary fluctuation of the
field $X_i$ around this background can be  expanded in terms $n\times
n$-matrix valued functions, the Weyl map generalises to,
\begin{equation}\label{gWeyl}
  \phi(x)=\sigma^\alpha \sqrt{\det\theta}\int \frac{\dd^p
  k}{(2\pi)^{p/2}}\e^{\ii kx} \tr (\sigma^\alpha\otimes \e^{\ii
  k\times p})\cdot \phi,
\end{equation}
where $\sigma^\alpha$ are generators of su$(n)$ with the normalisation
given by the $n\times n$-matrix trace,
\begin{equation}
  \tr_{(n)}\sigma^\alpha\sigma^\beta=\delta^{\alpha\beta}.
\end{equation}

\item Beyond the basic set of fields in \eqref{action} one can also
try to add some ``mater'' fields in the fundamental
representation. Let us note, however, that any real field in the
background invariant representation corresponds to a Hermitian
operator which realises the adjoint representation of the Hilbert
space unitary symmetry. In the star form such fields belong to adjoint
representation of the gauge group.  When trying to add complex fields
in the fundamental representation of the gauge group, one finds only
fields in the bi-fundamental representation of $G\times G$ where $G$
is the gauge group. Indeed, the complex field $\phi$ in the
fundamental representation should satisfy $\pd_\mu \phi(x)=\ii
p_\mu*\phi(x)$. Therefore, its action in the background form should
look like follows,
\begin{equation}\label{fund}
  S_{\mathrm{fund}}=\tr \left(\frac{1}{2}X_\mu\phi\phi^\dag
  X_\mu-V(\phi*\phi^\dag)\right).
\end{equation}
Obviously, beyond the desired symmetry,
\begin{equation}\label{fund-sym}
  \phi\to U^{-1}\phi,\qquad \phi^\dag\to \phi^\dag U,\qquad U\in G,
\end{equation}
there is another one
\begin{equation}\label{fund-sym2}
  \phi\to \phi V,\qquad \phi^\dag\to V^{-1}\phi^\dag,\qquad V\in G.
\end{equation}
\end{itemize}

\section{The Hilbert Space Picture}

As we have seen in the previous section, the model \eqref{action} may
look like a Yang--Mills model with scalar multiplet in various
dimensions or with various gauge groups depending on the background
solution chosen.

The common origin of these models leads to duality relations among
them. The roots of this duality are as follows. As we know algebras of
functions on different non-commutative spaces and with different gauge
groups are all isomorphic to the algebra of operators on the infinite
dimensional separable Hilbert space and, therefore isomorphic among
themselves. In particular, they are Morita equivalent. What is
important is that this isomorphism can relate smooth functions to
smooth i.e. preserves the topology of the algebra of functions. This
is in contrast to what one has in the commutative case.

In this section we are going to consider maps relating such models. We
will consider maps relating models with different gauge groups and
maps between models in different dimensions. In the next section we
consider even more particular case of map changing only the
non-commutativity parameter $\theta^{\mu\nu}$.

Before going to particular cases let us discuss some general aspects
of the map.

Consider two different backgrounds $p^{(1)}_{\mu_1}$ and
$p^{(2)}_{\mu_2}$ each having the form
\begin{equation}\label{pp}
  p^{(l)}_{\mu_l}=p'_{\mu_l}\otimes\I_{n_l},
\end{equation}
where $p'_l$, satisfy,
\begin{equation}
  [p'_{\mu_l},p'_{\nu_l}]=-\ii \theta^{-1}_{(l)\mu_l \nu_l}, \quad
  \det\theta_{(l)}\neq 0,
\end{equation}
and form a complete irreducible set of operators  on $\hh'_{(l)}$,
$l=1,2$.

The ranges $I(\mu_1)$ and $I(\mu_2)$ of indices $\mu_1$ and $\mu_2$
are two subsets of orders respectively $p_{(1)}$ and $p_{(2)}$ of the
sequence $1,\dots,D$.

Operators $p^{(l)}$, as given by eq. \eqref{pp} fail to form a
complete irreducible set due to their degeneracies when $n_l>1$.
These degeneracies are solved according to the previous section by
sets of $n_l\times n_l$-dimensional Pauli matrices
$\sigma^{(l)}_{\alpha_l}$, $\alpha_l=1,\dots,n_{(l)}^2-1$, which
together with $\sigma_0$ generate the algebra u$(n_l)$.
Correspondingly, the Hilbert space is split into,
\begin{equation}\label{hs-split}
  \hh\sim \hh'_{(l)}\otimes V_{(l)}, \qquad l=1,2.
\end{equation}

Now, using the definitions for the Weyl map and inverse Weyl map
\eqref{gWeyl} and \eqref{invWeyl}, one can write down the formula for
passing from one background to another. Thus, for a non-commutative
function $\phi_{(1)}(x_{(1)})$ with respect to the first background
one has a unique image $\phi_{(2)}(x_{(2)})$ with respect to the
second one, which is given by the following covariant map,
\begin{multline}\label{tr1}
  \phi_{(2)}(x_{(2)})=\\
  \sigma^{(2)}_{\alpha_2}\sqrt{\det\theta_{(2)}}\int\frac{\dd
  k_{(2)}}{(2\pi)^{p_{(2)}/2}}\e^{\ii k_{(2)}x_{(2)}}\int \dd x_{(1)}
  \tr_{V_{(1)}}E^{\alpha_2}(k^{(2)};x_{(1)})\phi_{(1)}(x_{(1)})\\
  \equiv \left(\SW\phi_{(1)}\right) (x_{(2)}),
\end{multline}
where,
\begin{equation}\label{tr2}
  E^{\alpha_2}(k_{(2)};x_{(1)})=\sigma^{(1)}_{\alpha_1} \int\frac{\dd
  k_{(1)}}{(2\pi)^{p_{(1)}}}\e^{\ii k_{(1)}x_{(1)}}\tr \left(
  \sigma^{(1)}_{\alpha_1}\otimes\e^{\ii k_{(1)}\times p_{(1)}}
  \sigma^{(2)}_{\alpha_2}\otimes\e^{\ii k_{(2)}\times p_{(2)}}\right)
\end{equation}
$\tr_{V_{(l)}}$ is the trace taken over $V_{(l)}$ indices. Trace with
no label is performed over the Hilbert space $\hh$.

This map defines transformation rules for all fields but gauge
ones. The gauge field in different backgrounds, in fact, corresponds
to Weyl symbols of different operators,
\begin{equation}\label{gf}
  A^{(l)}_{\mu_l}=X_{\mu_l}-p^{(l)}_{\mu_l}.
\end{equation}
Therefore, beyond the ``covariant'' part of transformation given by
formula analogous to eq. \eqref{tr1} there is also an inhomogeneous
part.

Also, due to the fact that some indices which in the first background
correspond to the gauge field in the second background may correspond
to the scalar field and vice versa if $\mu_1$ and $\mu_2$ run
different ranges ($I(\mu_1)\neq I(\mu_2))$), the map may interchange
the gauge field the scalar field.

Taking this into account one has for the map of the gauge field,
\begin{subequations}\label{trA}
\begin{align}
  &A^{(2)}_\alpha=\SW(p^{(1)}_\alpha+A^{(1)}_{\alpha})-
  p^{(2)}_\alpha,\qquad &\alpha\in I(\mu_2)\cap I(\mu_1),\\
  &A^{(2)}_\alpha=\SW\Phi^{(1)}_\alpha-p^{(2)}_\alpha, \qquad
  &\alpha\in I(\mu_2)\cap \bar{I}(\mu_1),\\
  &\Phi^{(2)}_\alpha=\SW(p^{(1)}_\alpha+A^{(1)}_{\alpha}), \qquad
  &\alpha\in \bar{I}(\mu_2)\cap I(\mu_1),\\
  &\Phi^{(2)}_\alpha=\SW\Phi^{(1)}_\alpha, \qquad &\alpha\in
  \bar{I}(\mu_2)\cap \bar{I}(\mu_1),
\end{align}
\end{subequations}
where $\SW$ is given by eq. \eqref{tr1}. The notations in the above
equations are as follows, $\alpha\in I(\mu_1)\cap I(\mu_2)$ means that
index $\alpha$ belongs to both range of $\mu_1$ and ones of $\mu_2$,
$\alpha\in I(\mu_1)\cap\bar{I}(\mu_2)$ means that $\alpha$ belongs to
the range of $\mu_1$ but not to one of $\mu_2$, and so on. In the
equations above we have not written the explicit dependence on
$x_{(l)}$, but assume that fields with $(l)$ label depend on
$x_{(l)}$, where $l=1,2$.

It is not difficult to verify that, under this map, gauge equivalent
configurations are mapped into gauge equivalent ones. In particular,
one has for the gauge fields,
\begin{equation}\label{SWmap}
  \left\{ \begin{array}{c} g_{(2)}^{-1}(p_{(2)}+A_{(2)})g_{(2)}\\
  g_{(2)}^{-1}\Phi_{(2)}g_{(2)} \end{array} \right\}=\SW\left\{
  \begin{array}{c} g_{(1)}^{-1}(p_{(1)}+A_{(1)})g_{(1)}\\
  g_{(1)}^{-1}\Phi_{(1)}g_{(1)} \end{array} \right\},
\end{equation}
where, $(A_{(1)},\Phi_{(1)})$ map into $(A_{(2)},\Phi_{(2)})$
according to \eqref{trA} while $g_{(1)}$ maps into $g_{(2)}$ according
to \eqref{tr1}. This means that the map we have obtained is a
\emph{Seiberg--Witten\/} map.

In the following subsections we consider the particular examples
realising either U(1)--U(2) duality or duality between models in
different dimensions.

\subsection{U(1) -- U($n$) duality}
Let us present the explicit construction for the map from U(1) to U(2)
gauge model in the case of two-dimensional non-commutative space. The
map we are going to discuss can be straightforwardly generalised to
the case of arbitrary even dimensions as well as to the case of
arbitrary U(n) group.

The two-dimensional non-commutative coordinates are,
\begin{equation}\label{2d}
  [x^1,x^2]=\ii \theta.
\end{equation}

The non-commutative analog of complex coordinates is given by
oscillator rising and lowering operators,
\begin{gather}\label{aabar}
  a=\sqrt{\frac{1}{2\theta}}(x^1+\ii x^2),\qquad
  \bar{a}=\sqrt{\frac{1}{2\theta}}(x^1-\ii x^2)\\
  a\ket{n}=\sqrt{n}\ket{n-1},\qquad \bar{a}\ket{n}=\sqrt{n+1}\ket{n+1},
\end{gather}
where $\ket{n}$ is the so-called oscillator basis formed by
eigenvectors of $N=\bar{a}a$,
\begin{equation}\label{n}
  N\ket{n}=n\ket{n}.
\end{equation}
The gauge symmetry in this background is non-commutative U(1).

We will now construct the non-commutative U(2) gauge model. For this,
consider the U(2) basis which is given by following vectors,
\begin{gather}\label{hv}
  \ket{n',a}=\ket{n'}\otimes e_a, \qquad a=0,1\\ e_0= \begin{pmatrix}
  1\\0 \end{pmatrix},\qquad e_1= \begin{pmatrix} 0\\1 \end{pmatrix},
\end{gather}
where $\{\ket{n'}\}$ is the oscillator basis and $\{e_a\}$ is the
``isotopic'' space basis.

The one-to-one correspondence between U(1) and U(2) bases can be
established in the following way \cite{Nair:2001rt},
\begin{equation}\label{2->1}
  \ket{n'}\otimes e_a\sim \ket{n}=\ket{2n'+a},
\end{equation}
where $\ket{n}$ is a basis element of the U(1)-Hilbert space and
$\ket{n'}\otimes e_a$ is a basis element of the Hilbert space of
U(2)-theory. (Note, that they are two bases of the same Hilbert space.)

Let us note that the identification \eqref{2->1} is not unique.  For
example, one can put an arbitrary unitary matrix in front of $\ket{n}$
in the r.h.s. of \eqref{2->1}. This in fact describes all possible
identifications and respectively maps from U(1) to U(2) model.

Under this map, the U(2) valued functions can be represented as scalar
functions in U(1) theory. For example, constant U(2) matrices are
mapped to particular functions in U(1) space. To find these functions,
it suffices to find the map of the basis of the u$(2)$ algebra given
by Pauli matrices $\sigma_\alpha$, $\alpha=0,1,\dots,3$.

In the U$(1)$ basis Pauli matrices look as follows,
\begin{subequations}\label{sigma}
 \begin{align} &\sigma_0=\sum_{n=0}^{\infty}
  \bigl(\ket{2n}\bra{2n}+\ket{2n+1}\bra{2n+1}\bigr)\equiv\I,\\
  &\sigma_1=\sum_{n=0}^{\infty} \bigl(\ket{2n}\bra{2n+1}+
  \ket{2n+1}\bra{2n}\bigr),\\ &\sigma_2=-\ii\sum_{n=0}^{\infty}
  \bigl(\ket{2n}\bra{2n+1}- \ket{2n+1}\bra{2n}\bigr),\\
  &\sigma_3=\sum_{n=0}^{\infty} \bigl(\ket{2n}\bra{2n}-
  \ket{2n+1}\bra{2n+1}\bigr), \end{align}
\end{subequations}
while the ``complex'' coordinates $a'$ and $\bar{a}'$ of the U(2)
invariant space are given by the following,
\begin{subequations}\label{aprim}
\begin{align}
  &a'=\sum_{n=0}^{\infty}\sqrt{n}\bigl(\ket{2n-2}\bra{2n}+
  \ket{2n-1}\bra{2n+1}\bigr),\\
  &\bar{a}'=\sum_{n=0}^{\infty}\sqrt{n+1}\bigl(\ket{2n+2}\bra{2n}
  +\ket{2n+3}\bra{2n+1}\bigr).
\end{align}
\end{subequations}

One can see that when trying to find the Weyl symbols for operators
given by \eqref{sigma}, \eqref{aprim}, one faces the problem that the
integrals defining the Weyl symbols diverge.  This happens because the
respective functions (operators) do not belong to the non-commutative
analog of $L^2$ space (are not square-trace).

Let us give an alternative way to compute the functions corresponding
to operators \eqref{sigma} and \eqref{aprim}. To do this let us
observe that operators
\begin{equation}\label{Pi+}
  \Pi_+=\sum_{n=0}^{\infty}\ket{2n}\bra{2n},
\end{equation}
and
\begin{equation}\label{Pi-}
  \Pi_-=\sum_{n=0}^{\infty}\ket{2n+1}\bra{2n+1},
\end{equation}
can be expressed as\footnote{Weyl symbols of $a$ and $\bar{a}$ are
denoted, respectively, as $z$ and $\bar{z}$. The same rule applies
also to primed variables.}
\begin{equation}\label{P+aa}
  \Pi_+=\frac{1}{2}\sum_{n=0}^{\infty}\left(1+
  \sin\pi\left(n+\frac{1}{2}\right)\right)\ket{n}\bra{n}\to
  \frac{1}{2}\left(1+\sin_*\pi\left(\bar{z}*z+\frac{1}{2}\right)\right),
\end{equation}
and,
\begin{equation}\label{P-aa}
  \Pi_-=\I-\Pi_+=
  \frac{1}{2}\left(1-\sin_*\pi\left(\bar{z}*z+\frac{1}{2}\right)\right)=
  \frac{1}{2}\left(1-\sin_*\pi|z|^2\right),
\end{equation}
where $\sin_*$ is the ``star'' sin function defined by the star Taylor
series,
\begin{equation}\label{star-sin}
  \sin_* f=f-\frac{1}{3!}f*f*f+\frac{1}{5!}f*f*f*f*f-\cdots,
\end{equation}
with the star product defined in variables $z,\bar{z}$ as follows,
\begin{equation}\label{aa*}
  f*g(\bar{a},a)=\e^{\pd\bar{\pd}'-\bar{\pd}\pd'}f(\bar{z},z)
  g(\bar{z}',z')|_{z'=z},
\end{equation}
where $\pd=\pd/\pd z$, $\bar{\pd}=\pd/\pd\bar{z}$ and analogously for
primed $z'$ and $\bar{z}'$. For convenience we denoted Weyl symbols of
$a$ and $\bar{a}$ as $z$ and $\bar{z}$.

The easiest way to compute \eqref{P+aa} and \eqref{P-aa} is to find
the Weyl symbol of the operator,
\begin{equation}\label{I}
  I^{\pm}_k=\frac{1\pm\sin\left(\bar{a}a+
  \frac{1}{2}\right)}{(\bar{a}a+\gamma)^k},
\end{equation}
were $\gamma$ is some constant, mainly $\pm 1/2$.

For sufficiently large $k$, the operator $I^\pm_k$ becomes square
trace for which the formula \eqref{Weyl} defining the Weyl map is
applicable. The Weyl symbol for smaller values of $k$ can be obtained
using the following recurrence relation,
\begin{equation}\label{rec}
  I^{\pm}_{k-m}(\bar{z},z)=
  \underbrace{\left(|z|^2+\gamma-\frac{1}{2}\right)*\dots*
  \left(|z|^2+\gamma-\frac{1}{2}\right)}_{m\text{ times}}
  *I^{\pm}_{k}(\bar{z},z).
\end{equation}
The last equation requires computation of only finite number of
derivatives of $I^\pm_{k}(\bar{z},z)$ arising from the star product
with polynomials in $\bar{z},z$.

{}From the viewpoint of the gauge theory \eqref{action}, the
configuration \eqref{aprim} can be seen as a solution to equations of
motion in the U(1) theory, while operators \eqref{sigma} are the ones
commuting with this solution, i.e.  generators of its symmetry algebra.

\subsection{Duality between models in different dimensions}

Consider the situation when $V$ is another Hilbert space or product of
Hilbert spaces. This topic was considered in
\cite{Sochichiu:2000bg,Sochichiu:2000kz} here we only shortly review
this.

Consider the Hilbert space $\hh$ corresponding to two-dimensional
non-commutative space \eqref{2d}, and $\hh\otimes\hh$ which
corresponds to four-dimensional non-commutative space generated by
\begin{equation}\label{4d}
  [x^1,x^2]=\ii\theta_{(1)},\qquad [x^3,x^4]=\ii\theta_{(2)}.
\end{equation}
In the last case non-commutative complex coordinates correspond to two
sets of oscillator operators,  $a_1$, $a_2$ and $\bar{a}_1$,
$\bar{a}_2$, where,
\begin{subequations}\label{4d-osc}
\begin{align}
  &a_1=\sqrt{\frac{1}{2\theta_{(1)}}}(x^1+\ii x^2),\qquad
  &\bar{a}_1=\sqrt{\frac{1}{2\theta_{(1)}}}(x^1-\ii x^2)\\
  &a_1\ket{n_1}=\sqrt{n_1}\ket{n_1-1},\qquad
  &\bar{a}_1\ket{n_1}=\sqrt{n_1+1}\ket{n_1+1},\\
  &a_2=\sqrt{\frac{1}{2\theta_{(2)}}}(x^3+\ii x^4),\qquad
  &\bar{a}_2=\sqrt{\frac{1}{2\theta_{(2)}}}(x^3-\ii x^4)\\
  &a\ket{n}_2=\sqrt{n_2}\ket{n_2-1},\qquad
  &\bar{a}_2\ket{n_2}=\sqrt{n_2+1}\ket{n_2+1},
\end{align}
\end{subequations}
and the basis elements of the ``four-dimensional'' Hilbert space
$\hh\otimes\hh$ are $\ket{n_1,n_2}=\ket{n_1}\otimes\ket{n_2}$.

The isomorphic map $\sigma:\hh\otimes\hh\to\hh$ is given by assigning
a unique number $n$ to each element $\ket{n_1,n_2}$ and putting it
into correspondence to $\ket{n}\in\hh$
\cite{Sochichiu:2000bg,Sochichiu:2000kz}.

As we discussed earlier, this map induces an isomorphic map of
connections and non-commutative functions from two to four dimensional
non-commutative spaces.

This can be easily generalised to the case with arbitrary number of
factors $\hh\otimes\dots\otimes\hh$ corresponding to $p/2$
``two-dimensional'' non-commutative spaces. In this way, one obtains
the isomorphism $\sigma$ which relates two-dimensional non-commutative
function algebra with a $p$-dimensional one, for $p$ even.

It is worthwhile to note that in the case of the map which relates
different dimensions, the flat connection is \emph{never\/} mapped to
flat one, since the tensor $\theta_{\mu\nu}$ is different in
$\hh_{(2)}$ and $\hh_{(2)}\otimes\hh_{(2)}$. (Obviously in the first
case it is two-dimensional while in the second one it is
four-dimensional.) This map again solves the Seiberg--Witten condition
that smooth bounded functions are mapped to smooth and bounded and
gauge equivalent configurations are mapped to gauge equivalent ones,
and therefore it is a non-perturbative Seiberg--Witten map. This map
is singular when $\theta= 0$.

In fact, the arguments above indicate that the model \eqref{action} is
non-perturbatively independent on $\theta_{\mu\nu}$ or the
dimensionality of the non-commutative Yang--Mills model. This fact is
called background independence \cite{Seiberg:2000zk}, also a
presumable feature of string field theory \cite{Witten:1992qy}.

\section{Perturbative Seiberg--Witten Map}
\setcounter{equation}{0}

So far, we have considered maps which relate algebras of
non-commutative functions in different dimensions or at least taking
values in different Lie algebras.  Due to the fact that they change
considerably the geometry, these maps could not be deformed smoothly
into the unit map. (At least it is not obvious that it can be done.)
In this section we consider a more restricted class of maps which do
not change either dimensionality or the gauge group but only the
non-commutativity parameter. Obviously, this can be smoothly deformed
into identity map, therefore one may consider infinitesimal
transformations.

In the approach of the second section the non-commutativity parameter
is given by the solution to the equations of motion. In this
framework, the SW map is given by the change of the background
solution $p_\mu$ to a slightly different one $p_\mu+\delta
p_\mu$. Then, a solution with the constant field strength $F^{(\delta
p)}_{\mu\nu}$ will change the non-commutativity parameter as follows,
\begin{equation}\label{delta-theta}
  \theta^{\mu\nu}+\delta
  \theta^{\mu\nu}\equiv(\theta^{-1}_{\mu\nu}+\delta\theta^{-1}_{\mu\nu})^{-1}=
  (\theta^{-1}_{\mu\nu}+F_{\mu\nu})^{-1}.
\end{equation}
Note, that the above equation does not require $\delta\theta$ to be
infinitesimal.

Since we are considering solutions to the gauge field equations of
motion $A_\mu=\delta p_\mu$ one should fix the gauge for it. A
convenient choice would be e.g. the Lorentz gauge, $\pd_\mu \delta
p_\mu=0$. Then, the solution with
\begin{equation}\label{delta-p}
  A^{(\delta p)}_\mu\equiv\delta p_\mu =(1/2)
  \epsilon_{\mu\nu}\theta^{\nu\alpha}p_\alpha
\end{equation}
with antisymmetric $\epsilon_{\mu\nu}$ has the constant field strength
\begin{equation}\label{F-delta}
 F^{(\delta p)}_{\mu\nu}\equiv
 \delta\theta^{-1}_{\mu\nu}=\epsilon_{\mu\nu}+(1/4)\epsilon_{\mu\alpha}
 \theta^{\alpha\beta}\epsilon_{\beta\nu}=\epsilon_{\mu\nu}+O(\epsilon^2).
\end{equation}
This corresponds to the following variation of the non-commutativity
parameter,
\begin{equation}\label{theta-new}
  \delta \theta^{\mu\nu}=-\theta^{\mu\alpha}\epsilon_{\alpha\beta}
  \theta^{\beta\nu}-\frac{1}{4}\theta^{\mu\alpha}\epsilon_{\alpha\gamma}
  \theta^{\gamma\rho}\epsilon_{\rho\beta}\theta^{\beta\nu}=
  -\theta^{\mu\alpha}\delta\theta^{-1}_{\alpha\beta}
  \theta^{\beta\nu}+O(\epsilon^2).
\end{equation}
Let us note that such kind of infinitesimal transformations were
considered in a slightly different context in \cite{Ishikawa:2001mq}.

Let us find how non-commutative functions are changed with respect to
this transformation. In order to do this, let us consider how the Weyl
symbol \eqref{Weyl} transforms under the variation of background
\eqref{delta-p}.  For an arbitrary operator $\phi$ after short
calculation we have,
\begin{equation}\label{delta-Phi}
  \delta \phi (x)=\frac{1}{4} \delta\theta^{\alpha\beta}(\pd_\alpha
  \phi*p_\beta(x)+p_\beta*\pd_\alpha \phi (x)).
\end{equation}
In obtaining this equation we had to take into consideration variation
of $p_\mu$ and of the factor $\sqrt{\det\theta}$ in the definition of
the Weyl symbol \eqref{Weyl}.

By construction, this variation satisfies the  ``star-Leibnitz rule'',
\begin{equation}\label{prop1}
  \delta (\phi*\chi)(x)=\delta
  \phi*\chi(x)+\phi*\delta\chi(x)+\phi(\delta*)\chi(x),
\end{equation}
where $\delta\phi(x)$ and $\delta\chi(x)$ are defined according to
\eqref{delta-Phi} and variation of the star-product is given by,
\begin{equation}\label{delta-*}
  \phi(\delta*)\chi(x)=\frac{1}{2}\delta\theta^{\alpha\beta}
  \pd_\alpha\phi*\pd_\beta\chi(x).
\end{equation}
The property \eqref{prop1} implies that $\delta$ provides an
homomorphism (which is in fact an isomorphism) of star algebras of
functions.

The above transformation \eqref{delta-Phi} do not apply, however, to
the gauge field $A_\mu(x)$ and gauge field strength
$F_{\mu\nu}(x)$. This is the case because the respective operators are
not background independent.  Indeed, according to \eqref{base-sol}
$A_\mu=X_\mu-p_\mu$, where $X_\mu$ is background
independent. Therefore the gauge field $A_\mu(x)$ transforms
inhomogeneously,
\begin{equation}\label{delta-A}
  \delta A_\mu(x)=\frac{1}{4}\delta\theta^{\alpha\beta} (\pd_\alpha
  A_\mu*p_\beta+p_\beta*\pd_\alpha A_\mu)+
  \frac{1}{2}\theta_{\mu\alpha}\delta\theta^{\alpha\beta}p_\beta.
\end{equation}

The transformation law for $F_{\mu\nu}(x)$ can be computed using
\eqref{F} and the ``star-Leibnitz rule'' \eqref{prop1} as well as the
fact that it is the Weyl symbol of the operator,
\begin{equation}\label{xx->F}
  F_{\mu\nu}=\ii[X_\mu,X_\nu]-\theta_{\mu\nu}.
\end{equation}
Of course, both approaches give the same result,
\begin{equation}\label{delta-F}
  \delta F_{\mu\nu}(x)=\frac{1}{4}\delta\theta^{\alpha\beta}
  (\pd_\alpha F_{\mu\nu}*p_\beta+p_\beta*\pd_\alpha
  F_{\mu\nu})(x)-\delta\theta^{-1}_{\mu\nu}.
\end{equation}

The infinitesimal map described above has the following properties:
\begin{enumerate}
  \item[$i$).] It maps gauge equivalent configurations to gauge
equivalent ones, therefore it satisfies the Seiberg--Witten equation,
\begin{equation}\label{SWE}
  U^{-1}*A*U+U^{-1}*dU\to U^{\prime-1}*'A'*'U'+U^{\prime-1}*'d'U'.
\end{equation}
  \item[$ii$).] It is linear in the fields.  \item[$iii$).] Any
  background independent functional is invariant under  this
  transformation. In particular, any gauge invariant functional whose
  dependence on gauge fields enters through the combination
  $X_{\mu\nu}(x)=F_{\mu\nu}+\theta^{-1}_{\mu\nu}$ is invariant with
  respect to \eqref{delta-Phi}--\eqref{delta-F}. This is also the
  symmetry of the action provided that the gauge coupling transforms
  according to \eqref{g}.  \item[$iv$).] Formally, the transformation
  \eqref{delta-Phi} can be represented in the form,
\begin{equation}\label{xf}
  \delta \phi(x)=\delta x^\alpha \pd_\alpha \phi (x)=\phi (x+\delta
  x)-\phi(x),
\end{equation}
where $\delta x^\alpha=-\theta^{\alpha\beta}\delta p_\beta$ and no
star product is assumed. This looks exactly like a coordinate
transformation.
\end{enumerate}

One may naturally raise the question: how is this connected with the
``standard'' SW map found in \cite{Seiberg:1999vs}?

In \eqref{delta-p} we have chosen $\delta p_\mu$ independent of gauge
field background. (In fact the gauge field background was switched-on
later, after the transformation.) An alternative way would be to have
nontrivial field $A_\mu(x)$ from the very beginning and to chose
$\delta p_\mu$ to be of the form,
\begin{equation}\label{d-pSW}
  \delta_{\mathrm{SW}}p_\mu=-\frac{1}{2}\epsilon_{\mu\nu}
  \theta^{\nu\alpha}A_\alpha.
\end{equation}
Then, the transformation laws corresponding to such a transformation
of the background coincide exactly with the standard SW map. The
substitution \eqref{d-pSW} is possible because the function
$p_\mu=-\theta^{-1}_{\mu\nu}x^\nu$ has the same gauge transformation
properties as $-A_\mu(x)$,
\begin{equation}\label{upu}
  p_\mu\to U^{-1}*p_\mu*U(x)-U^{-1}*\pd_\mu U(x).
\end{equation}

\section{The Quantum Theory}
\setcounter{equation}{0}

In the previous section we found that the map \eqref{delta-p} changes
the fields leaving the gauge invariant functionals including the
action unchanged.  To be a symmetry in the ``quantum field theory''
e.g. the path integral formulation of ``quantum'' theory,  one should
check the invariance of the quantum measure as well.

Consider the partition function corresponding to the model
\eqref{p-dim}. In the traditional approach, the path integral
representation for the partition function is obtained via canonical
quantisation. This approach can be easily generalised also to models
with spatial non-commutativity. In the case of non-degenerate
space-time non-commutativity one can use path integral as the
definition for the partition function and, generally, of the
non-commutative ``quantum'' theory.

So, consider the path integral representation for the partition
function
\begin{equation}\label{Z}
  Z=\int [\dd A][\dd \Phi]\det M \e^{\ii S+\ii S_{\mathrm{g.f.}}},
\end{equation}
where $[\dd A]$ and $[\dd \Phi]$ are usual functional measures for
Weyl symbols, $S_{\mathrm{g.f.}}$ is the gauge fixing term and $\det
M$ Faddeev--Popov determinant. For example generalised Lorentz gauge
correspond to the choice,
\begin{equation}\label{Lorentz}
  S_{\mathrm{g.f.}}^{\mathrm{Lorentz}}=\int  \frac{1}{2\alpha}(\pd_\mu
  A_\mu)^2,
\end{equation}
and Faddeev--Popov determinant,
\begin{equation}\label{fpM}
  \det M=\det \pd_\mu \nabla_\mu.
\end{equation}

Let us analyse the background invariance properties of the quantum
theory as given by the  formal path integral \eqref{Z}.  As
established in the previous section, the classical action is invariant
under changes of the background \eqref{delta-p} provided respective
rules for transformation of fields, products and couplings are
applied. What remains to be established to generalise the
``classical'' results is the invariance of the functional measure.

Another source of trouble would be the gauge fixing term which was
added to the theory during ``quantisation'' and which may spoil the
background invariance. Apparently, it is difficult to find a gauge
fixing term which would be background invariant. This happens because
background invariant functionals are all gauge invariant. However, we
claim that the way the gauge fixing term destroys gauge invariance is
an in-offensive one, and  indicates only that the gauge fixing
prescription does depend on the background.  Moreover at finite volume
(finite Hilbert space) it is not necessary.

To establish the transformation properties of the measure, let us
recall that the measure can be seen as the determinant of the
invariant functional quadratic form,
\begin{equation}\label{norm}
  \|\bar{\delta}A_\mu\|^2=\int \dd x\, (\bar{\delta}A_\mu(x))^2,\qquad
  \|\bar{\delta}\Phi\|^2=\int \dd x\, (\bar{\delta}\Phi_a(x))^2,
\end{equation}
where $\bar{\delta}A_\mu(x)$ and $\bar{\delta}\Phi_a(x)$ are
independent variations of field $A_\mu(x)$ and $\Phi_a(x)$. The
equations \eqref{norm} can be rewritten as follows,
\begin{equation}\label{Hilbert-norm}
  \|\bar{\delta}A_\mu\|^2=
  (2\pi)^{p/2}\sqrt{\det\theta}\tr(\bar{\delta}X_\mu)^2,\qquad
  \|\bar{\delta}\Phi\|^2=(2\pi)^{p/2}\sqrt{\det\theta}(\bar{\delta}\Phi_a)^2,
\end{equation}
where in the last equations, the Hilbert space operators are used and
$\bar{\delta}A_\mu$ is replaced by the equivalent $\bar{\delta}X_\mu$.

As one can see from the equations \eqref{Hilbert-norm}, the dependence
of functional norm from the background enters only through factors
$(2\pi)^{p/2}\sqrt{\det\theta}$. (The first factor $(2\pi)^{p/2}$ is
important only in the case when dimensionality changes, which is not
the case for infinitesimal background transformations.) Therefore,
variation of norms with respect to change of background
\eqref{delta-p} is as follows,
\begin{subequations}\label{delta-norm}
\begin{align}
  &\delta \|\bar{\delta}A_\mu\|^2=\frac{1}{2}\delta
  (\ln\det\theta)\|\bar{\delta}A_\mu\|^2=
  -\frac{1}{2}\theta^{-1}_{\alpha\beta}
  \delta\theta^{\alpha\beta}\|\bar{\delta}A_\mu\|^2\\ &\delta
  \|\bar{\delta}\Phi_a\|^2=\frac{1}{2}\delta
  (\ln\det\theta)\|\bar{\delta}\Phi_a\|^2=
  -\frac{1}{2}\theta^{-1}_{\alpha\beta}
  \delta\theta^{\alpha\beta}\|\bar{\delta}\Phi_a\|^2.
\end{align}
\end{subequations}

According, to Eqs. \eqref{delta-norm} the functional measure changes
as follows,
\begin{subequations}\label{delta-measure}
\begin{align}
  &\delta [\dd A]=(p
  \theta^{-1}_{\alpha\beta}\delta\theta^{\alpha\beta} \int\dd x) [\dd
  A],\\ &\delta [\dd \Phi]=((D-p)\theta^{-1}_{\alpha\beta}\delta
  \theta^{\alpha\beta}\int\dd x) [\dd \Phi],\\ &\delta [\dd A][\dd
  \Phi]=(D \theta^{-1}_{\alpha\beta}\delta \theta^{\alpha\beta}\int\dd
  x) [\dd A][\dd\Phi].
\end{align}
\end{subequations}
Thus, we end up with an ``anomaly'',
\begin{equation}\label{anomaly}
  \frac{\delta Z}{\delta\theta^{\alpha\beta}}=(D
  \theta^{-1}_{\alpha\beta} \V)Z,
\end{equation}
where $\V$ is the (regularised) volume of space-time.

Since the ``anomaly'', we obtained in such a way is a constant
proportional to the the volume of the space-time,  it can be absorbed
in a ($\theta$-dependent) renormalisation  of the vacuum energy.

Let us note, that the ``anomaly'' is proportional to the factor $D$
which is the number of bosonic fields. Fermions, while transforming in
the same way as the bosonic fields under the Seiberg--Witten map, the
fermionic measure contributes with an opposite sign.  Then, in
supersymmetric models, or at least in models with equal numbers of
bosons and fermions the ``anomaly'' \eqref{anomaly} vanishes.

\subsection{Remarks on Regularisation}

All these results were obtained in the naive approach when no
regularisation and renormalisation is taken into consideration. The
regularisation and renormalisation may change drastically some
conclusions concerning the fate of the classical symmetries  of the
theory and we will discuss this issue here.

For the case of pure non-commutative Yang--Mills--scalar model one can
write down regularisation schemes satisfying all necessary
criteria. Higher covariant derivative or dimensional regularisation
schemes seem to satisfy the background invariance.  A particularly
interesting regularisation of the non-commutative Yang--Mills model
would be the finite $N$ IKKT matrix model \cite{Ishibashi:1996xs}
obtained by the truncation of the Hilbert space to $N$
dimensions. This regularisation is non-perturbative, beyond this it is
suitable for the numeric computations. The required properties follow
from the explicit background invariant form of the action.

Although, the algebra \eqref{irred} for finite $N$ is altered this
model approaches the non-commutative Yang--Mills model in the
background invariant form \eqref{action} in the limit
$N\to\infty$. This method is good enough if we deal with a purely
bosonic theory. For fermion containing models there can appear
problems connected with fermionic spectrum doubling
\cite{Sochichiu:2000fs}. Another potential problem of this method can
be identified with the existence of non-compact directions in the path
integral associated with flat directions in the action.  For IKKT-type
matrix models these are associated with flat directions in the
potential. Such potential infinities have been investigated
\cite{Krauth:1998yu}, \cite{Ambjorn:2000bf} both at finite N. It turns
out that for sufficiently large N (typically larger than four) the
measure of the path integral is convergent.  Thus, finite-N truncation
is a valid background invariant non-perturbative regularisation.

In continuum perturbation theory a momentum cutoff regularisation is
the popular choice. For finite cutoff $\Lambda$ IR-UV mixing implies
the existence of IR singularities \cite{Minwalla:1999px} reflecting
the UV divergences of the commutative theory. No renormalisation
procedure is yet known which deals with such singularities. Such
singularities are logarithmic at one-loop in supersymmetric theories
and as such, amenable to standard renormalisation. However, it was
shown that power singularities appear at higher loops, and their
resummation seems not possible \cite{kosh}. It has been argued
\cite{VanRaamsdonk:2001jd} that such power IR singularities reflect
linear potentials among constituent $D_0$ branes in a matrix model
realisation of non-commutative theories indicating an instability for
non-supersymmetric theories.

It is thus fair  to say that in perturbation theory, a momentum cutoff
regularisation and renormalisation of non-commutative theories is an
open problem.

The use of finite-N truncation of the Hilbert space in the
perturbative expansion is not popular. One reason is that it is not
convenient in the continuum formulation of perturbation theory. Also,
it seems to modify the ``classical background" around which we do
perturbation theory: the commutation relations $[x^i,x^j]=\theta^{ij}$
with constant $\theta$ cannot be satisfied in a finite dimensional
Hilbert space. We do believe however, that their slight modification
by a projection operator, at large N, where N is the dimension of the
Hilbert space is innocuous for the regularisation although it can
contribute to finite quantities after renormalisation. From the other
hand, it allows non-perturbative analysis in which the system may
visibly choose the ``preferred'' background.

Combined with the aforementioned result, that Hilbert space truncation
is a regularisation at large N, it seems to be the right scheme in
order to discuss the fate of the classical dualities we have presented.

We have shown above that up to a constant renormalisation of the path
integral, classical duality is a symmetry of the path integral and the
unregulated measure. Thus, any ``anomaly" to this symmetries will
appear as a non-invariance of the cutoff.

In the finite-N truncation regularisation scheme it is straightforward
to answer this question. We will consider as a motivating example the
simplest classical map between a U(1) and an U($n$) non-commutative
gauge theory with one non-commutative plane. Let us define in the U(1)
case the regularised theory by truncating to the first $nN$ states of
the harmonic oscillator Hilbert space. The Hilbert space of the U($n$)
gauge theory is the tensor product of a $n$ dimensional vector space
and the harmonic oscillator Hilbert space. We regularise the theory by
keeping the first $N$ states of this Hilbert space. Then the classical
map is one-to-one on the remaining finite vector spaces. This map
introduces the correspondence between the cutoff parameters of these
theories. Thus, the cutoff $\Lambda$ in U(1) theory corresponds to the
cutoff of order $\Lambda/n^{2/p}$ in the U($n$) model.

This type of cutoff respects the duality symmetry. Similar cutoff
procedures exist for the other duality maps. It is obvious however
that singularities are expected if one considers the large $n$-limit
of non-commutative U($n$) gauge theory.

In general the cutoffs in different backgrounds can be related via
background invariant quantities. Thus, in the large $N$ limit one has
background invariant trace,
\begin{equation}\label{cutoff1}
  N=\tr \I=\frac{n}{(2\pi)^{p/2}\sqrt{\det\theta}}\int_R\dd^p x
  =\frac{n R^p}{(2\pi)^{p/2}\sqrt{\det\theta}},
\end{equation}
where $R$ is the IR cutoff in the theory. This implies that under the
equivalence maps,
\begin{equation}\label{cutoff2}
  \frac{n R^p}{(2\pi)^{p/2}\sqrt{\det\theta}}= \frac{n
  \sqrt{\det\theta}\Lambda^p}{(2\pi)^{p/2}}=\text{invariant},
\end{equation}
where $\Lambda$ is the UV cutoff and we also used the explicit
relation between the IR and UV cutoffs, $R^p=(\det{\theta})\Lambda^p$,
which follows, e.g. from the fact that $x^\mu=-\theta^{\mu\nu}p_\nu$.

Perturbation theory in a given dual version breaks explicitly the
duality invariance since different perturbation theories correspond to
expansions around different backgrounds of the universal theory. It is
thus, not surprising that different perturbation theories have
different physics. What we claim here is that there is a
non-perturbative definition of the theory which accommodates the
classical duality symmetries with no anomalies.  This heavily relies
on the renormalisation procedure (that so far is not well understood)
respecting the symmetry. It is in principle possible however  that the
symmetry be broken spontaneously due to dynamics. Whether this is
realized is beyond our tools at the moment.

It seems that were it for the duality to survive a properly
regularised theory, it would imply non-perturbative equaivalence of
models in different number of non-commutative dimensions.  In
perturbation theory such theories are plagued by (generically) power
IR singularities that render the physics of the theory obcure.
Quantum validity of duality will imply that such divergences are
artifacts of perurbation theory.  They should not be there in a
non-pperturbative treatement.  On the other hand, resummation of
subsectors of perturbation theory are not expected to lead to
substancial iprovement.

\section{Conclusion}

In the present work we have studied dualities in the non-commutative
gauge models. Classically, such dualities relate gauge models in
different dimensions as well as models with different ``local'' gauge
groups. ``Smaller'' transformations relate the same model on spaces
with different non-commutativity parameter $\theta$. This is similar
to the situation with the Seiberg--Witten map.  Indeed, the map
satisfies the condition that it maps gauge equivalent configurations
to gauge equivalent ones and satisfies an appropriate differential
equation.

 Nevertheless, in the case of ``small'' maps when one can compare our
solution with the Seiberg--Witten ansatz it appears to be
different. The difference consists in the fact that our solution is
linear in fields and has a different structure of singularities.

The above classical duality symmetries can be extended to the quantum
theory. In the path integral approach this is equivalent to the
invariance of the measure. Naively, i.e. neglecting the issues with
IR/UV divergences, the functional measure is always duality invariant.
For the consistency one should regularise and renormalise the
theory. In the purely bosonic theory it is possible to present such a
regularisation, and respective renormalisation.  In the case of chiral
fermions, however, problems can appear. In general since the duality
symmetry is intrinsically connected with the gauge invariance we
believe that gauge anomaly-free theories should possess also
background invariance at the quantum level.

In the opposite case, quantum breaking of this symmetry (an anomaly)
may  not be fatal for the consistency of the theory.  It may signal
the appearance of inequivalent perturbative vacua.
Non-perturbatively, the theory may favour some of these vacua relative
to others.

Another interesting feature of the maps described here is that they
relate models with different couplings and different cutoffs.  This
may reveal interesting information about various regimes of the gauge
models. Unfortunately, so far there is no reliable description of the
Quantum Non-perturbative Field theories even in the weak coupling
regime due to problems related with IR/UV mixing.

It is an important open problem to understand better a background
invariant renormalisation of non-commutative theories. Even in
perturbation theory (which breaks background invariance) such a
procedure is not understood.  This will send light to the role duality
maps play in the physics of non-commutativity.

\acknowledgments

We would like to thank A. Koshelev and K. Anagnostopoulos for many
discussions. The work of C.S. was partially supported by RFBR grant,
Scientific School Support grant No. 00-15-96046, and INTAS grant
No. 00-262 E.K. was partially supported by RTN contracts
HPRN--CT--2000-00122 and --00131 and INTAS contract   N 99 1 590.
\providecommand{\href}[2]{#2}\begingroup\raggedright
\endgroup
\end{document}